\documentclass[10pt]{article}
\usepackage[OE]{express}
\usepackage{siunitx}
\usepackage{subfigure}
\usepackage{booktabs}
\usepackage{multirow}

\begin{document}

\title{Optical manipulation for studies of collisional dynamics of micron-sized droplets under gravity}

\author{Maksym Ivanov,\authormark{1} Kelken Chang,\authormark{2} Ivan Galinskiy,\authormark{3} Bernhard Mehlig,\authormark{2} and Dag Hanstorp\authormark{2,*}}

\address{\authormark{1}Vilnius University Laser Research Center, LT-10223 Vilnius, Lithuania \\
\authormark{2}Department of Physics, University of Gothenburg, SE-412 96 Gothenburg, Sweden \\
\authormark{3}Facultad de Ciencias, Departamento de Fisica, Universidad Nacional Autonoma de Mexico, Ciudad Universitaria, CdMx 04510, Mexico}

\email{\authormark{*}dag.hanstorp@physics.gu.se}



\begin{abstract}
A new experimental technique for creating and imaging collisions of micron-sized droplets settling under gravity is presented.
A pair of glycerol droplets is suspended in air by means of two optical traps.
The droplet relative velocities are determined by the droplet sizes.
The impact parameter is precisely controlled by positioning the droplets using the two optical traps.
The droplets are released by turning off the trapping light using electro-optical modulators.
The motion of the sedimenting droplets is then captured by two synchronized high-speed cameras, at a frame rate of up to 63~kHz.
The method allows the direct imaging of the collision of droplets without the influence of the optical confinement imposed by the trapping force.
The method will facilitate efficient studies of the microphysics of neutral, as well as charged, liquid droplets and their interactions with light, electric field and thermodynamic environment, such as temperature or vapor concentration.
\end{abstract}



\section{Introduction}
\label{sec:Intro}

In their textbook on the microphysics of clouds, Pruppacher and Klett \cite{pruppacher:2010} devote an entire chapter to collisions of droplets settling under gravity.
The primary concern in this problem is the determination of the collision efficiency.
The collision efficiency, defined as the ratio of the collision cross-section to the geometric cross-section, is a dimensionless measure of the probability of collision.
In addition to determining the droplet growth rates (see e.g. \cite{pruppacher:2010}), the collision efficiency is a key ingredient in the collision theory of cloud droplets widely used in many weather and climate models \cite{shaw:2003, grabowski:2013}.
The collision efficiency is the most uncertain aspect of collision rate theories \cite{grabowski:2013}.
Experimental determination of the collision efficiency remains elusive \cite{pruppacher:2010, low:1982, beard:1995, bordas:2013, szakall:2014, nagare:2015}, although recent studies have yielded some empirical conclusions \cite{devenish:2012}.

In recent years, advances in optical manipulation and high-speed imaging techniques have provided the opportunity of controlling and measuring the size and trajectories of individual droplets in air.
In a pioneering study, Ashkin and Dziedzic \cite{ashkin:1975} observed the collision and coalescence between an optically levitated droplet and a gravitationally settling droplet.
This study was followed by a large number of investigations.
For example, Hopkins \textit{et al.} \cite{hopkins:2004} used optical tweezers to observe the coagulation of aerosol droplets, and they determined the droplet growth rates using cavity-enhanced Raman spectroscopy (CERS) with nanometer precision.
Power \textit{et al.} \cite{power:2012} used a single-beam gradient-force optical trap to coalesce droplets of \SI{2}{\micro\metre} to \SI{12}{\micro\metre} in diameter and observed the elastically scattered light from the trapped particles to investigate the time-resolved dynamics of mixing.
Horstmann \textit{et al.} \cite{horstmann:2012} constructed a single-aerosol particle trap and coupled it to a conventional optical tweezers to guide and coalesce droplets of 500~nm to \SI{19}{\micro\metre} in diameter.
In all of the above studies, optical tweezers provide a fully controllable and convenient way to investigate droplet collisions.
However, the light field interacts with the trapped droplets and modifies their dynamics.
The work presented here demonstrates a new method of optical manipulation technique that circumvents this problem.
The method allows us to precisely image droplet collisions in free fall.

\section{Experimental background}
\label{sec:Background}

\begin{figure}[ht!]
\centering\includegraphics[height=5cm]{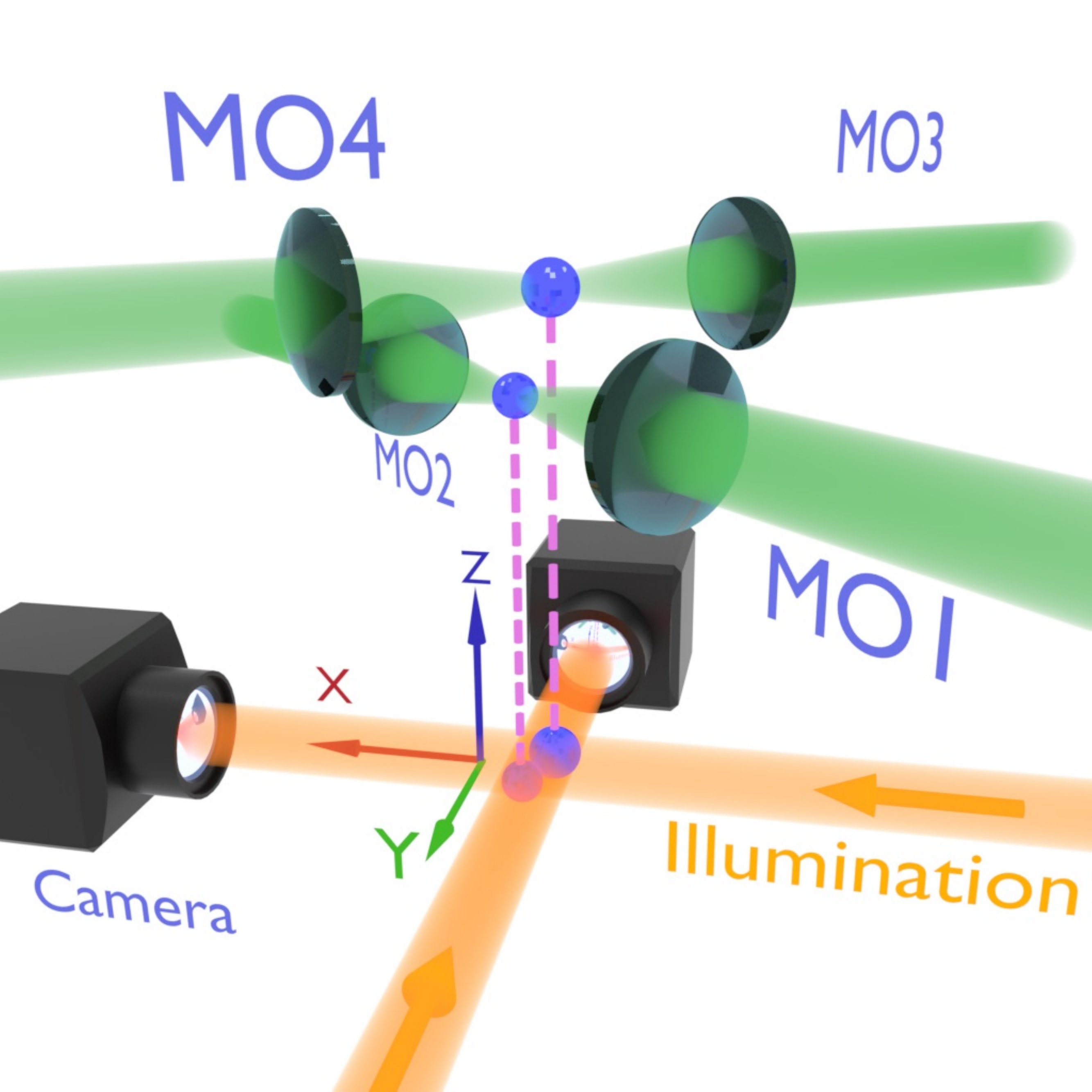}
\caption{The principle scheme for studies of droplet interactions. MO1, MO2 and MO3, MO4 are microscope objectives that focus the laser light and form the ``lower'' and ``upper'' optical trap, respectively.  The droplets are illuminated with LEDs, and their motion recorded by a pair of cameras with their lines of sight arranged at a \ang{90} angle.}
\label{fig:CollisionSchematic}
\end{figure}

We trap two glycerol droplets of different sizes in air by means of two pairs of focused counter-propagating beams (Fig. \ref{fig:CollisionSchematic}).
Two microscope objectives MO1 and MO2 focus light from two counter-propagating beams to a common focal point, forming the lower trap for the small glycerol droplet.
Similarly, the microscope objectives MO3 and MO4 focus light from two laser beams to a second point, forming the upper trap for the large glycerol droplet.
The two traps are vertically separated.
By switching off the laser beams with the appropriate timing, the droplets fall under the influence of gravity and may collide at a predictable distance below the lower trap.
The droplet motion is captured by two high-speed digital video cameras arranged in a horizontal plane.
Their lines of sight intersect at an angle of \ang{90}.

\begin{figure}[ht!]
\begin{center}
\subfigure[]{\includegraphics[height=5.25cm]{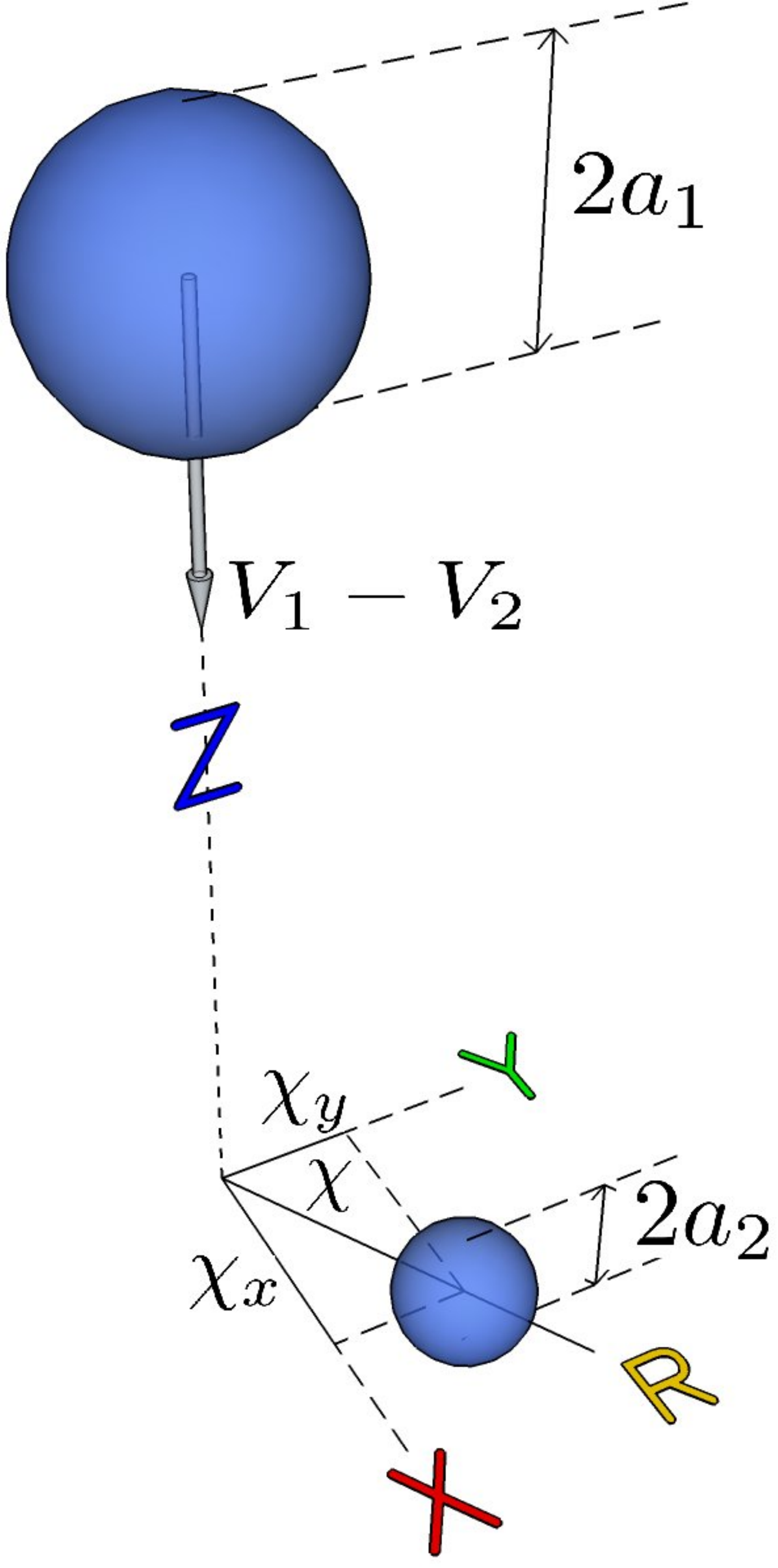}} \hspace{2cm}
\subfigure[]{\includegraphics[height=5.25cm]{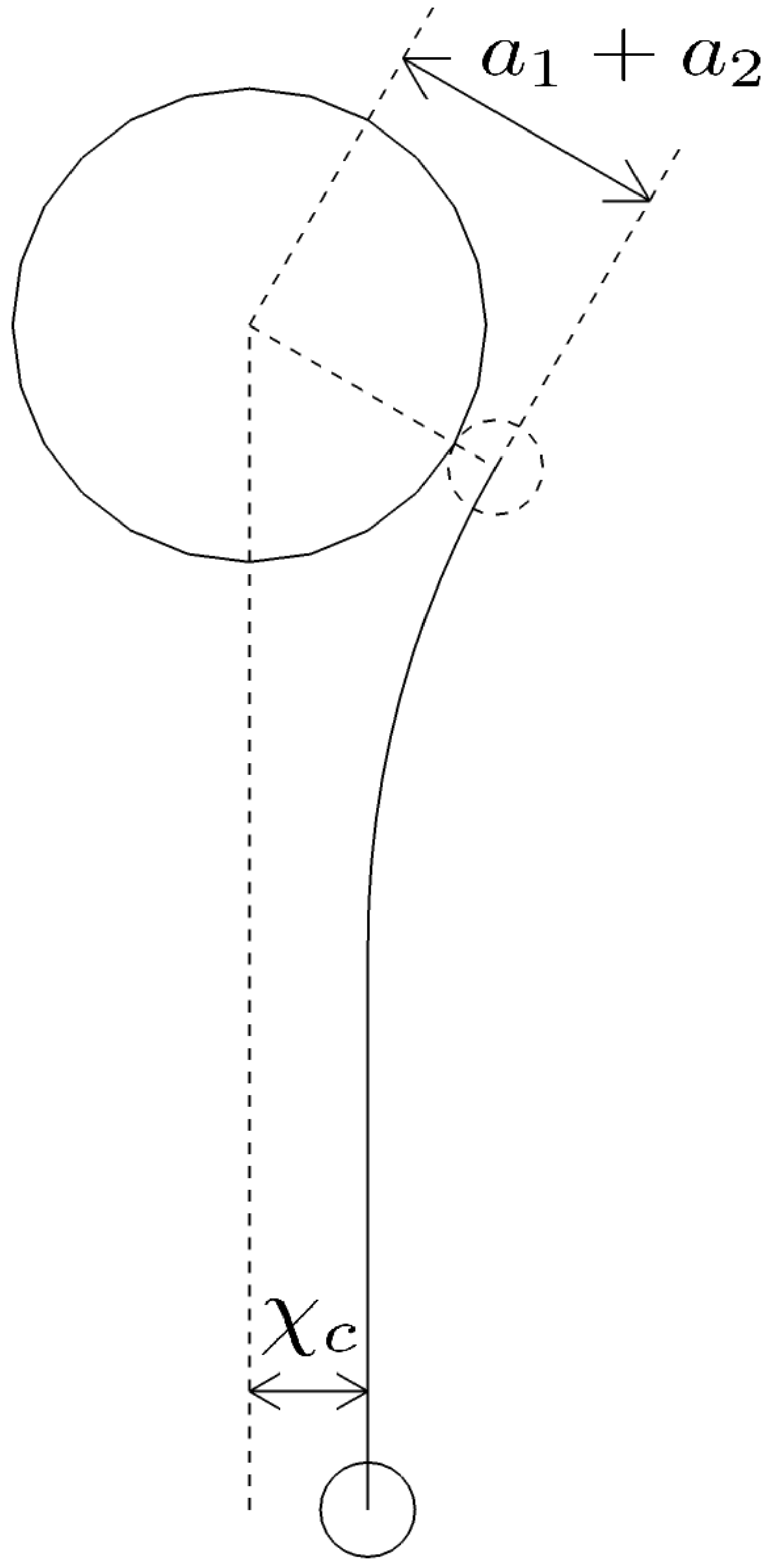}}
\caption{Panel (a) shows the geometry for the determination of the impact parameter, $\chi$, of the collision process from its two orthogonal projections on $x$ and $y$-axes. Two droplets with radii $a_1$ (top) and $a_2$ (bottom) fall in gravitational field, along the $z$-axis. The parameters $\chi_x$ and $\chi_y$ are the projections of the impact parameter $\chi$ onto the $x$ and $y$-axes, respectively. Panel (b) shows the trajectory of the interacting droplets under gravity, as observed from the reference frame of the larger droplet. $\chi_c$ is the critical impact parameter for a grazing trajectory of the smaller droplet.}
\label{fig:ImpactParameter}
\end{center}
\end{figure}
The geometry of the collision process is described in Fig. \ref{fig:ImpactParameter}.
A droplet with radius $a_1$ is positioned above a second droplet with radius $a_2$, where $a_1 > a_2$.
The impact parameter, $\chi$, is the projection of the distance between the centers of the two droplets perpendicular to their initial relative velocity, $V_1 - V_2$.
The projections of the impact parameter onto the $x$-$z$ and $y$-$z$ planes yield $\chi_x$ and $\chi_y$, respectively, and they are related by the equation
\begin{equation}
\label{eq:ImpactParameter}
\chi = \sqrt{\chi_x^2 + \chi_y^2} \,.
\end{equation}
In the experiment, we control the droplet impact parameter $\chi$ by adjusting the relative horizontal position between the two traps.
At a critical impact parameter, $\chi_c$, a grazing trajectory results, as depicted in Fig. \ref{fig:ImpactParameter}(b).
The collision efficiency is defined as $E = \chi_c^2 / (a_1 + a_2)^2$.
The geometrical arrangement of the droplets is described by the dimensionless impact number, defined as \cite{jiang:1992}
\begin{equation}
{\rm B} = \frac{\chi}{a_2 \, (1 + \Gamma)} \,,
\end{equation}
where $\chi$ is the impact parameter defined in equation (\ref{eq:ImpactParameter}), $a_2$ is the radius of the small droplet and $\Gamma = a_1 / a_2$ is the size ratio.
Thus, values of $0$ or $1$ for the impact number designate head-on or grazing collisions, respectively.

A droplet falling under the influence of gravity and Stokes drag in still air acquires a terminal velocity (Eq. (10-138) of \cite{pruppacher:2010})
\begin{equation}
V = \frac{2}{9} \, \bigg( \frac{\rho_d}{\rho_a} - 1 \bigg) \, \frac{g\, a^2}{\nu} \,,
\label{eq:TerminalVelocity}
\end{equation}
where $g = 9.81$~ms$^{-2}$ is the acceleration of gravity, $a$ is the droplet radius, $\nu = 1.6 \times 10^{-5}$m$^2$s$^{-1}$ is the kinematic viscosity of air, and $\rho_d = 1.26 \times 10^3$~kg m$^{-3}$ and $\rho_a = 1.23$~kg m$^{-3}$ are the densities of the droplet and air, respectively.
In our study, the droplets are composed of glycerol, but other types of chemical composition can also be used.
The Reynolds number
\begin{equation}
{\rm Re} = \frac{V \, a}{\nu} \,,
\end{equation}
measures the effect of convective forces relative to the fluid viscous forces \cite{gustavsson:2016}.
A value of Re $\ll 1$ allows a Stokesian description of the motion of the droplet.
In this process, there are two relevant time scales.
First, the droplet relaxes to its terminal velocity in time $\tau_d = V/g$.
Second, the representative convective time scale in quiescent air is $\tau_a = a / V$.
The ratio of these time scales defines the dimensionless Stokes number
\begin{equation}
{\rm St} = \frac{\tau_d}{\tau_a} = \frac{2}{9} \, \bigg( \frac{\rho_d}{\rho_a} - 1 \bigg) \, {\rm Re} \,.
\end{equation}
The Stokes number measures the importance of particle inertia.

Whether or not the droplet kinetic energy can overcome the interfacial energy barrier and coalesce is measured by the Weber number, defined in terms of the radius of the small droplet \cite{tang:2012}
\begin{equation}
{\rm We} = \frac{2 \, a_2 \, \rho_d \, (V_1 - V_2)^2}{\sigma} \,,
\end{equation}
where $\sigma = 6.34 \times 10^{-2}$~Nm$^{-1}$ is the surface tension of glycerol.
The droplet physical and kinematic parameters investigated in this work are summarized in table \ref{table:DropletParameters}.
\begin{table}[htbp]
\caption{The droplet physical characteristics and the collision parameters in the experiments.}
\begin{center}
\begin{tabular}{l*{6}{c}}
\toprule
\addlinespace[8pt]
Experimental section & \multicolumn{2}{c}{\ref{subsec:CoalescenceInTrap}} & \multicolumn{2}{c}{\ref{subsec:KissAndTumbling}} & \multicolumn{2}{c}{\ref{subsec:GravitationalCoalescence}} \\
\addlinespace[5pt]
\midrule
\addlinespace[8pt]
Diameter, $2a$~(\SI{}{\micro\metre}) & $31.9 $ & $40.0$ & $29.9$ & $37.9$ & $29.9$ & $33.9$ \\
\addlinespace[8pt]
Terminal velocity, $V$~(cms$^{-1}$) & 3.55 & 5.58 & 3.12 & 5.01 & 3.12 & 4.01 \\
\addlinespace[8pt]
Reynolds number, Re & 0.04 & 0.07 & 0.03 & 0.06 & 0.03 & 0.04 \\
\addlinespace[8pt]
Stokes number, St & 8.0 & 15.9 & 6.6 & 13.5 & 6.6 & 9.7 \\
\addlinespace[8pt]
Diameter ratio, $\Gamma$ & \multicolumn{2}{c}{1.25} & \multicolumn{2}{c}{1.27} & \multicolumn{2}{c}{1.13} \\
\addlinespace[8pt]
Impact number, B & \multicolumn{2}{c}{0.51} & \multicolumn{2}{c}{0.50} & \multicolumn{2}{c}{0.14} \\
\addlinespace[8pt]
Weber number, We & \multicolumn{2}{c}{$2.6 \times 10^{-4}$} & \multicolumn{2}{c}{$2.1 \times 10^{-4}$} & \multicolumn{2}{c}{$4.7 \times 10^{-5}$} \\
\addlinespace[5pt]
\bottomrule
\end{tabular}
\label{table:DropletParameters}
\end{center}
\end{table}

\section{The experiment}
\subsection{Optics and imaging system}
\label{subsec:Optics}

\begin{figure}[ht!]
\centering\includegraphics[width=9cm]{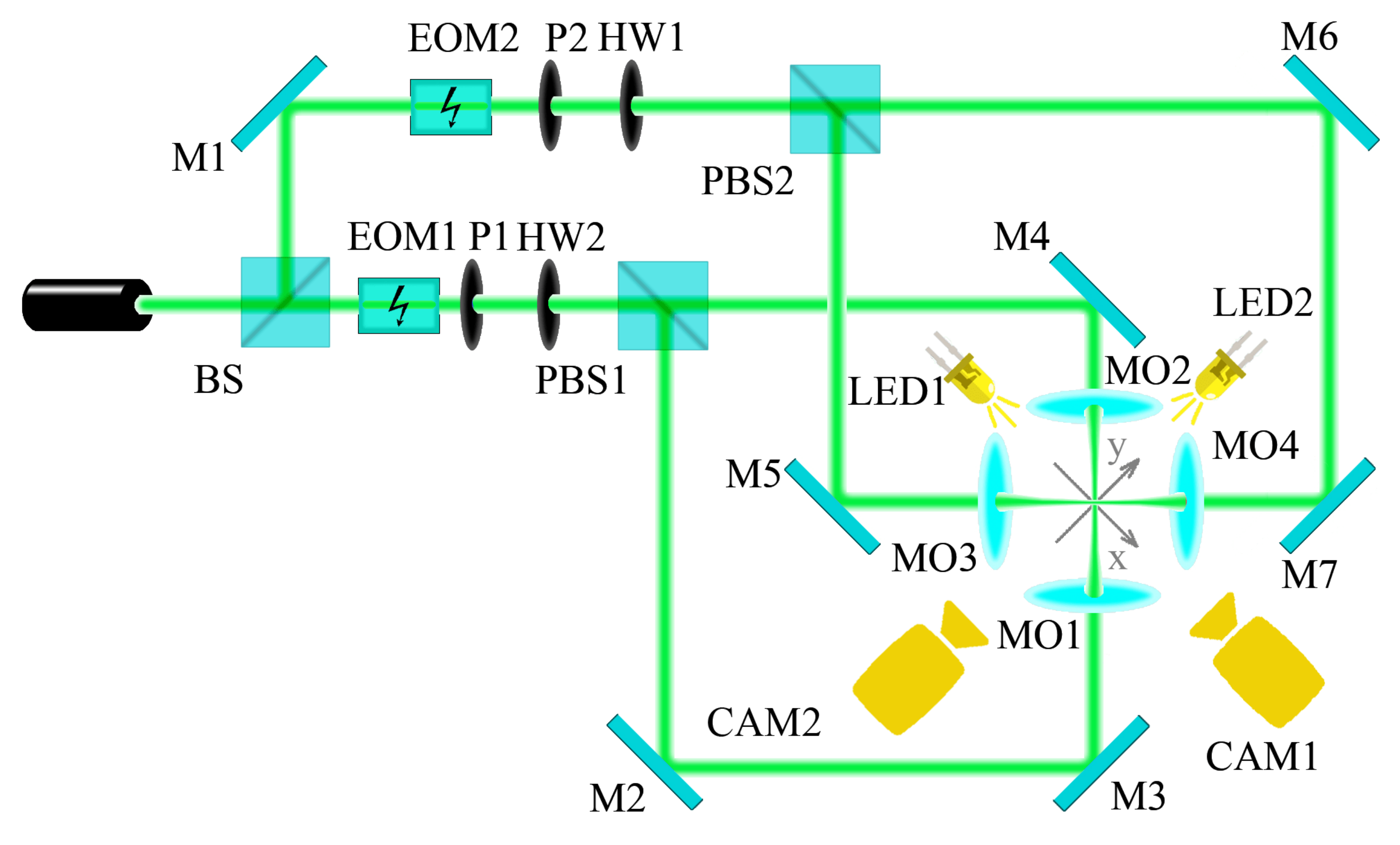}
\caption{Sketch of the experimental setup. BS - beam splitter, EOM - electro-optical modulators, M - mirrors, P - polarizers, HW - half-wave plates, PBS - polarizing beam splitters, MO - micro-lenses, LED - illumination, CAM - high speed digital movie camera. The $x$-$y$ coordinate system represents the laboratory frame. The $z$-axis points out of the page, whereas gravity points into the page.}
\label{fig:Optics}
\end{figure}

A sketch of the experimental setup is shown in Fig. \ref{fig:Optics}.
A laser beam generated by a solid state laser (Laser quantum ``gem532'', 532 nm, 2~W maximum power) passes through a beam splitter (BS) to form  two optical traps at different heights, which we named the ``upper'' and ``lower'' trap.
The transmitted beam that is collinear to the incident beam is used to form the ``lower'' trap.
This beam is split a second time by a polarizing beam splitter PBS1.
A micro-lens (MO) with focal length 5~mm focuses each beam such that the foci of the two counter-propagating beams overlap.
The combined trapping power from the two beams is approximately 350~mW.
In a similar manner, the reflected beam from BS is used to create the ``upper'' trap, which is positioned approximately 2~mm above the lower trap.
Two half wave plates (HW1 and HW2) and the polarizing beam splitters ensure that the counter propagating beams have orthogonal polarization states, thus eliminating undesirable interference at the focal points.
Two electro-optical amplitude modulators (EOM1 and EOM2) and two polarizers (P1 and P2) allow us to control the timing of the release of the droplets. The region where the droplets interact is illuminated by two collimated cold white light LED (Thorlabs MCWHL5) operated at 10\% of the maximum power.
Shadow images are observed by two perpendicularly mounted high speed cameras (Phantom Miro LAB310).
In order to minimize heating by the LED the illumination is pulsed synchronously with the droplet collisions.
The cameras are equipped with Infinity Model K2 DistaMax Long-Distance Microscope System set to $5\times$-$30\times$ magnification with a mean working distance of 13~cm.
Laser light from the trapped droplets is blocked by a notch filter (Thorlabs NF533-17), so that the cameras see only shadow images of the droplets with a bright Poisson spot in the center.

\subsection{Droplet generation and size control}
\label{subsec:DropletDispenser}

We generate droplets in the size range from 4 to \SI{60}{\micro\metre} in diameter using a commercial printer cartridge (Hewlett-Packard C6614).
The cartridge works as a drop-on-demand system, where a droplet is ejected each time a TTL pulse is sent to the cartridge from a pulse generator (see \cite{sergeyev:2006} and \cite{galinsky:2015} and references within for the droplet generation technique).
The cartridge, washed and filled with a solution mixed with 90\% (by volume) of distilled water and 10\% of glycerol (produced by Fisher Scientific with a purity of 99.6\%), produces droplets with a uniform diameter of approximately \SI{22}{\micro\metre}.
The water evaporates rapidly as the droplet descends down towards the trap. Hence, the size of the trapped droplet can be controlled by changing the mixing ratio of glycerol solution.
As an example, a solution containing 5\% of glycerol in water yields droplets of approximately \SI{11}{\micro\metre} in diameter after the water has completely evaporated away.
Alternatively, the droplet can be made larger by allowing two drops to coalesce in the trap.
The size of the largest droplet that can be levitated is determined  by the laser beam waist and the available laser power.
For the results presented in this work, the droplet size distribution is in the range from 29.9 to \SI{40.0}{\micro\metre} in diameter.

\subsection{The droplet chamber}
\label{subsec:Chamber}

An important part of the setup is the chamber for delivery, trapping and interaction of droplets.
Air current within the chamber easily affects the droplet motion.
In order to shield the droplets from air turbulence, the cartridge and micro-lenses are mounted in an enclosed chamber.
The chamber is made out of metal and cover glass plates connected by flexible construction sealant (Sikaflex 291i).
The flexible connection allows the movement of the micro-lenses to be decoupled from the metal plates of the chamber.
Hence, the position of the optical traps can be independently adjusted without breaking the seal.
Droplets descend from the cartridge through a flexible polymer tubing and enter the chamber through a glass cylindrical tube (2.5~cm in diameter) placed above the chamber.
Droplets ejected from the cartridge travel a few centimeters before reaching their terminal velocity and descend vertically down the tube to the trapping points.
The chamber has four windows on the sides for illumination and optical access and one on top for visual control and physical access.
The volume of the chamber is approximately $0.24~\si{liters}$.

\subsection{The initial control}
\label{subsec:InitialControl}

In order to produce a collision, the two droplets are released from the same horizontal position but at different vertical positions.
As a coarse adjustment the separation of the traps is adjusted by moving the lenses for the optical traps (MO1/MO2 or MO3/MO4) in pair.
Fine adjustment in the separation of the traps is accomplished by distributing the input power in one of the counter-propagating beams for the optical trap asymmetrically.
This is done with a slight rotation of the half wave plate HW1 or HW2 that is placed before the beam enters the polarizing beam splitters PBS1 or PBS2 (Fig. \ref{fig:Optics}).
The vertical separation of the two optical traps is 1.7~mm, which is sufficiently large for the upper droplet to reach its terminal velocity before approaching the lower droplet.

\subsection{The release mechanism}
\label{subsec:ReleaseMechanism}

The droplets are released from the optical traps by switching off the laser beams using electro-optical amplitude modulators (EOM1 and EOM2) (Thorlabs EO-AM-NR-C4) in combination with linear polarizers (P1 and P2) (Fig. \ref{fig:Optics}).
The polarizers P1 and P2 are aligned along the initial polarization vector of the laser  beam.
By applying the half wave voltage $V_\pi$ to the EOMs (in our case $V_{\pi} = 186$~V), the EOMs rotate the polarization by \ang{90}, so it is being blocked by the polarizers P1 and P2.
Hence, the optical traps are turned off and the droplets are released.

\subsection{The acquisition system and timing mechanism}
\label{subsec:AcquisitionAndTiming}

The imaging system consists of two high-speed digital movie cameras (Phantom Miro LAB310 from Vision Research) arranged in a horizontal plane with an angular separation of approximately \ang{90}.
Shadow images of the droplets are obtained  by projecting incoherent collimated light from two LEDs (Thorlabs MCWHL5) onto the sensors of the cameras (see e.g. \cite{nishino:2000}).
This arrangement not only offers a simple approach in the alignment of the two droplets, but also allows a precise determination of the droplet positions and impact parameters.
Movies of the droplet motion are recorded synchronously by the cameras at a resolution of $64 \times 768$ pixels (width $\times$ height).
The field of view is sufficiently large to observe both the trapped droplets and the interaction region.
In order to resolve the droplet motion at high spatial resolution, each camera is equipped with a K2 DistaMax long-distance microscope from Infinity Photo-Optical Company.
Each camera pixel observes an area of \SI{3.98}{\micro\metre} $\times$ \SI{3.98}{\micro\metre} in space, so that the total field of view is 0.26~mm by 3.06~mm (width $\times$ height).
The maximum frame rate at this resolution is 63000~Hz.
After loading the droplets into both traps, the LED illumination projects shadows of the droplets onto the camera sensors for the fine positioning (see section \ref{subsec:InitialControl}).
Both cameras and voltage supply for the EOMs are synchronised and controlled by an external pulse generator (BNC model 565).
At time $t_0$, the pulse generator delivers a trigger signal to the voltage supply of the EOM2 to release the upper droplet, and to initiate the recording of the movies on both cameras.
At time $t_{2/3}$ when the upper droplet has traveled $2/3$-rd of its way to the lower droplet, a second signal synchronized with the first one triggers the voltage supply to the EOM1 to release the lower droplet.
The cameras continue imaging the motion of both droplets until they are out of view.
For the initial separation of 1.7~mm, $2/3$-rd of it is $1.134$~mm and $t_{2/3}$ is approximately $100 \pm 50$~ms, depending on the droplet size.
For a glycerol droplet of \SI{31.9}{\micro\metre} in diameter, $t_{2/3}$ is 100.012~ms.

\subsection{Spatial resolution}
\label{subsec:SpatialResolution}

To map out the laboratory coordinates in real space, a Thorlabs calibration mask (model R2L2S3P1) containing uniform dots of \SI{62.5}{\micro\metre} in diameter arranged in a square lattice with a separation of $\ell =$\SI{125}{\micro\metre} between adjacent dots is placed in the mutual focal plane of the cameras.
An automated particle center finding routine, written in the Matlab programming language, was used to extract the two-dimensional coordinates of the center of the dots, from which the mean separation between adjacent dots $L$ (in pixels) was derived.
The spatial resolution of each pixel, $R$, was obtained from $R = \ell / L$.
In our experiments, the spatial resolution was approximately $4.0 \pm$\SI{0.4}{\micro\metre} per pixel.
Because the light sources were collimated, the error in determining the diameters of the droplets is expected to be fairly small, which is shown as follows.
The areas of the dot shadows were measured to size the dots.
The areas were equal to the number of pixels below a certain threshold.
By this method, the shadow diameters of the dots were consistently about 1.06 times larger than the actual diameter of the dots.
After correction for the bias, the uncertainty in determining the mean diameters was about \SI{2}{\micro\metre}, which is representative of the error in the sizing of the droplets.

\section{Results}
\label{sec:Results}
We have recorded 127 movies of collisions of droplets with different impact parameters and sizes.
In the following sections, we present three main cases: the coalescence of a gravitationally settling droplet with an optically trapped droplet (section \ref{subsec:CoalescenceInTrap}), the collision between two gravitationally settling droplets in the absence of trapping laser light resulting in non-coalescence \ref{subsec:KissAndTumbling}), and coalescence (section \ref{subsec:GravitationalCoalescence}).
In all three cases we show the trajectories of droplets as they fall, collide and possibly coalesce.

\subsection{Coalescence of droplets in an optical trap}
\label{subsec:CoalescenceInTrap}

\begin{figure}[ht!]
\centering\includegraphics[width=9cm]{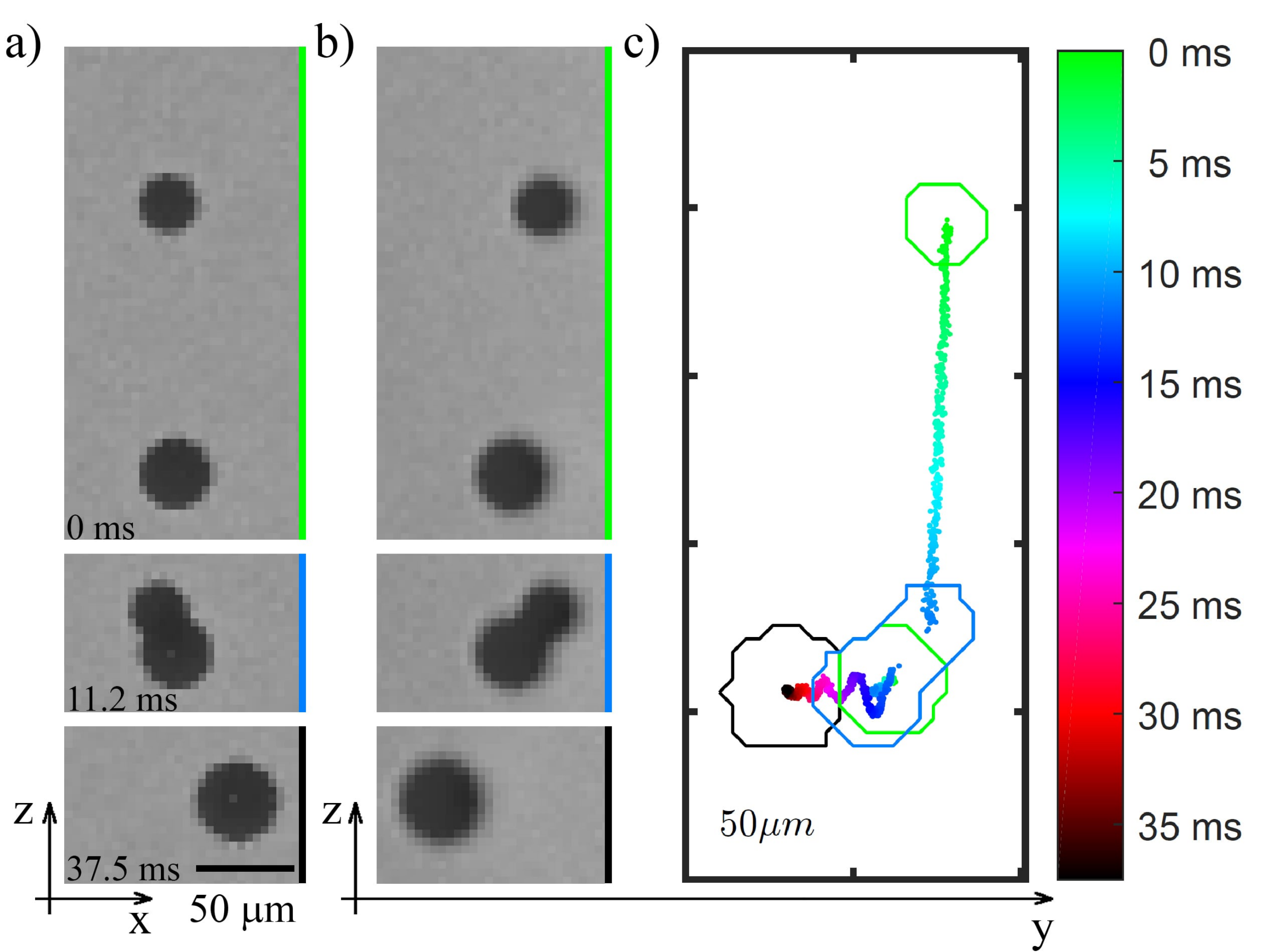}
\caption{Coalescence of two droplets while the bottom droplet is trapped by laser light as viewed in (a) the $x$-$z$ plane and (b) the $y$-$z$ plane. Panel (c) shows the droplet trajectories as well as their surfaces extracted from the movie recorded in panel (b). Adjacent tick marks on both axes indicate spatial separation of \SI{50}{\micro\metre}. The color bar indicates the progression in time in milliseconds.}
\label{fig:CoalescenceInTrap}
\end{figure}

Figure \ref{fig:CoalescenceInTrap} shows the coalescence process between a gravitationally settling droplet and an optically trapped droplet, imaged at a frame rate of 32~kHz, similar to the experiments in \cite{ashkin:1975,hopkins:2004,power:2012,horstmann:2012}.
The settling droplet is $31.9 \pm$\SI{2.0}{\micro\metre} in diameter, whereas the optically trapped droplet is $40.0\pm$\SI{2.0}{\micro\metre} in diameter, and the resultant droplet is $43.8\pm$\SI{2.0}{\micro\metre} in diameter.
The high value of impact number ${\rm B} = 0.51$, calculated from the projections of droplet separations $\chi_x$ and $\chi_y$, indicates no coalescence in the case of absence of laser field.
In this experiment, the laser light in the lower trap holds the larger droplet stationary in space and guides the settling droplet towards the trap, as can be seen in the trajectories in Fig. \ref{fig:CoalescenceInTrap}.
As the two droplets reach a critical separation, they merge to form a larger drop.
The resulting droplet stays in the trap and executes a damped oscillatory motion (see \textcolor{blue}{Visualization 1}).
For clarity, this video has been slowed down by a factor of 320 to 100~fps.

\begin{figure}[ht!]
\begin{center}
\subfigure[]{
\includegraphics[scale=0.27]{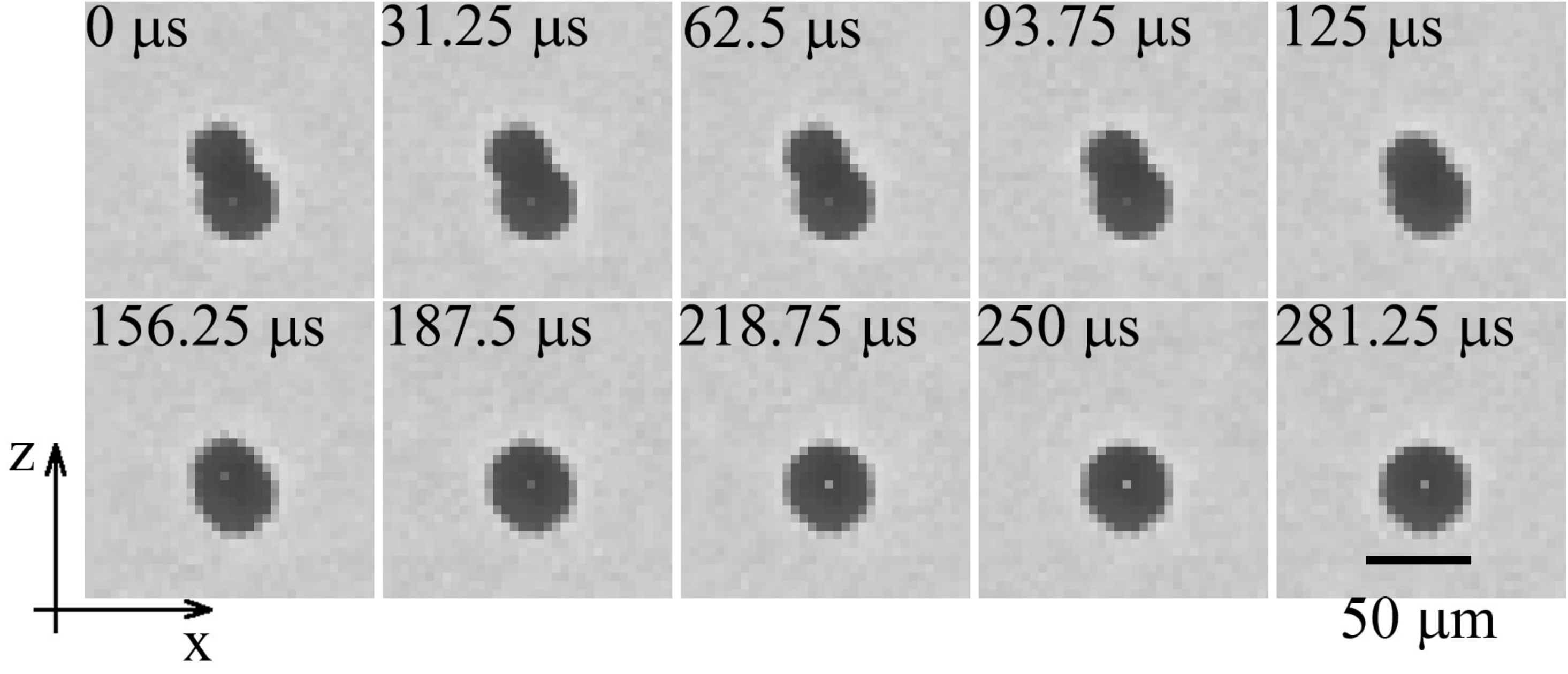}}
\subfigure[]{
\includegraphics[scale=0.27]{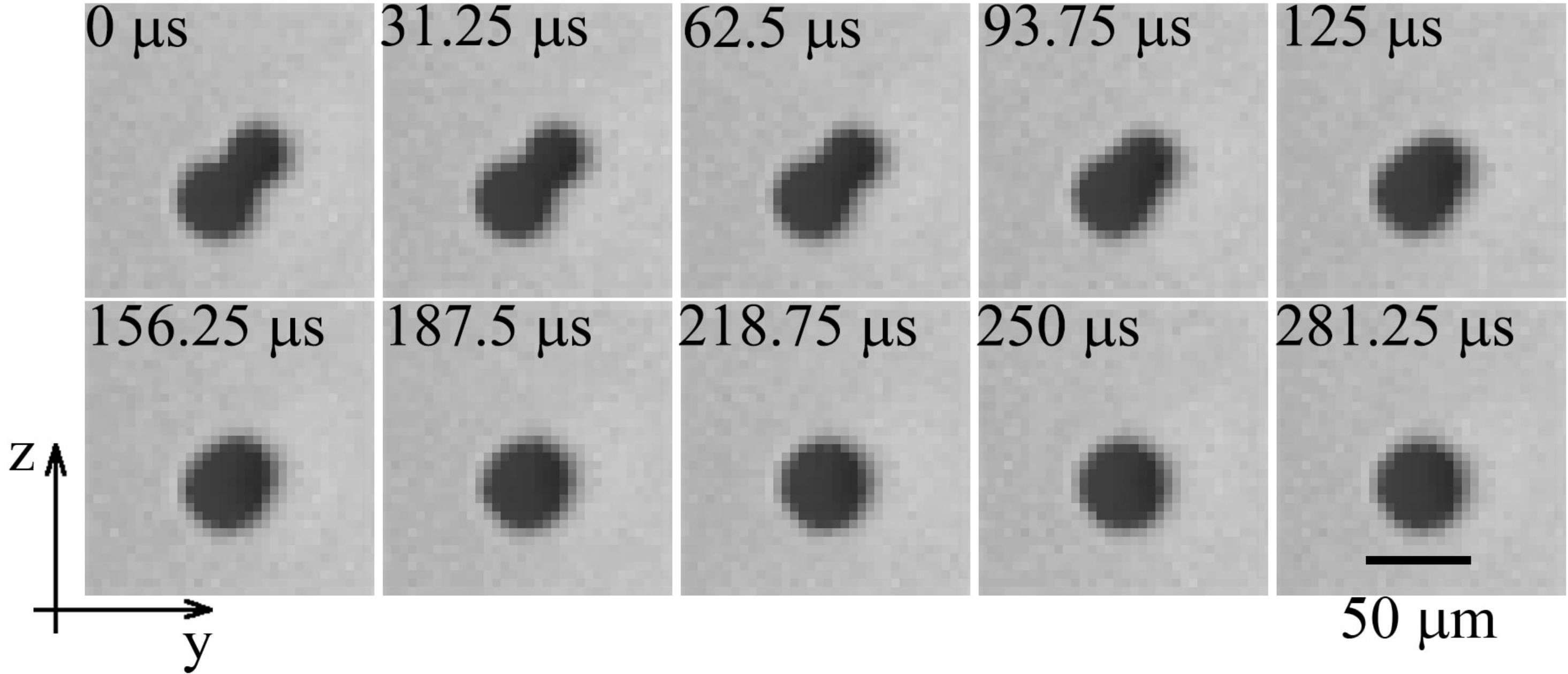}}
\caption{Projection on $x$ and $y$ directions of time resolved coalescence of two droplets when the bottom droplet is trapped by laser light.}
\label{fig:CoalescenceInTrapProjections}
\end{center}
\end{figure}

Figure \ref{fig:CoalescenceInTrapProjections} shows the coalescence process that is shown in Fig. \ref{fig:CoalescenceInTrap} in a more detailed frame-by-frame sequence of images at a temporal resolution of \SI{31.25}{\micro s} per frame.
In both sequences, the coalescence process is completed within a time scale of approximately \SI{180}{\micro s}.
In comparison with earlier works \cite{hopkins:2004,power:2012,horstmann:2012}, in which the droplets were guided by the trapping laser light, in our case droplet coalescence is still the favored outcome.

\subsection{Gravitational settling with kiss-and-tumbling motion}
\label{subsec:KissAndTumbling}

\begin{figure}[ht!]
\centering\includegraphics[width=8.5cm]{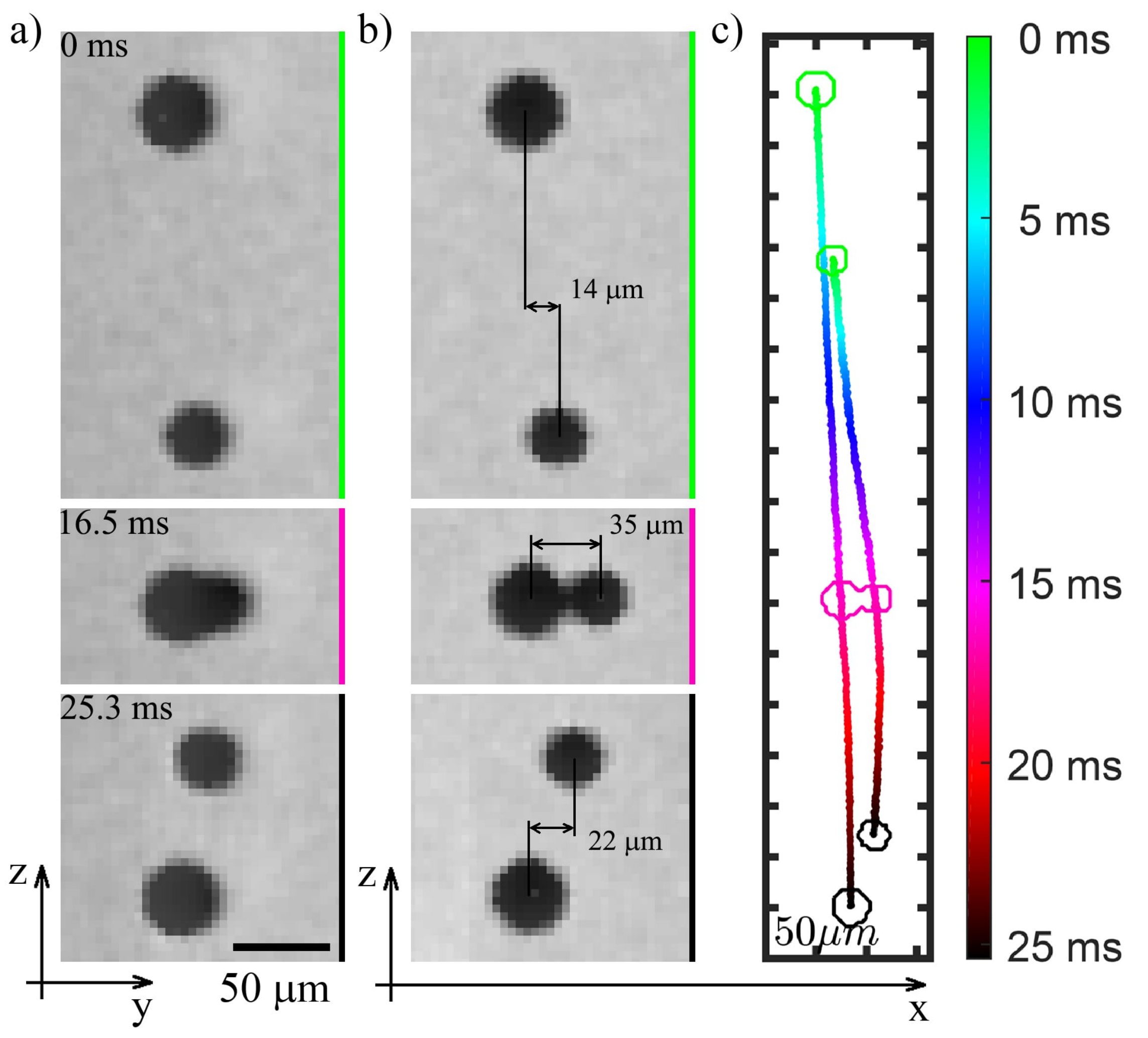}
\caption{The motion of two droplets settling in quiescent air under gravity as viewed on (a) the $y$-$z$ plane and (b) the $x$-$z$ plane that results in kiss-and-tumbling motion. Panel (c) shows the trajectories of the two droplets as seen in panel (b). Adjacent tick marks on both axes indicate spatial separation of \SI{50}{\micro\metre}. Color bar indicates temporal progression in milliseconds.}
\label{fig:KissAndTumbling}
\end{figure}

Figure \ref{fig:KissAndTumbling} shows the motion of two droplets of different sizes settling under gravity, imaged at a frame rate of 63~kHz.
The diameter of the larger droplet is $37.9\pm$\SI{2.0}{\micro\metre} and the diameter of the smaller droplet is $29.9\pm$\SI{2.0}{\micro\metre}.
The impact number, calculated from the projection of droplet separations $\chi_x$ and $\chi_y$, is ${\rm B} = 0.5$.
Before the larger droplet affects the smaller one (0~ms in Fig. \ref{fig:KissAndTumbling}) the separation between them is $\chi_x = 14 \pm$\SI{2}{\micro\metre} along the $x$-axis.
Consideration based on the initial geometry of the droplet trajectories predicts that they will collide.
However, as the large droplet approaches the small one, it deflects the small droplet to the side without touching, so much so that when the large droplet finally catches up with the small one the separation $\chi_x$ reaches a maximum of $35\pm$\SI{2.0}{\micro\metre} (at $t = 16.5$~ms on Fig. \ref{fig:KissAndTumbling}).
As the large droplet continues to fall, the small droplet moves around the large one and gradually recovers its initial vertical motion ($t = 25.3$~ms in Fig. \ref{fig:KissAndTumbling}) due to Stokesian microscopic reversibility.
This microscopic kiss-and-tumbling motion, whereby the small droplet bends its trajectory around the big droplet without touching, has been theoretically described by Zhang and Davis \cite{zhang:1991} and observed in experiment for sub-millimeter-size droplets \cite{zhang:1993}.
The total interaction time, which is approximately the duration of time the centers of the droplets stay within a separation of $a_1 + a_2$ from each other, is approximately $4.2 \pm 0.1$~ms.
The movie describing this process is available as supplementary material (see \textcolor{blue}{Visualization 2}).
The temporal resolution in this movie is \SI{15.86}{\micro s}.
For clarity, this movie has been slowed down to 100~fps.

\subsection{Gravitational settling with permanent coalescence}
\label{subsec:GravitationalCoalescence}

\begin{figure}[ht!]
\centering\includegraphics[width=8.5cm]{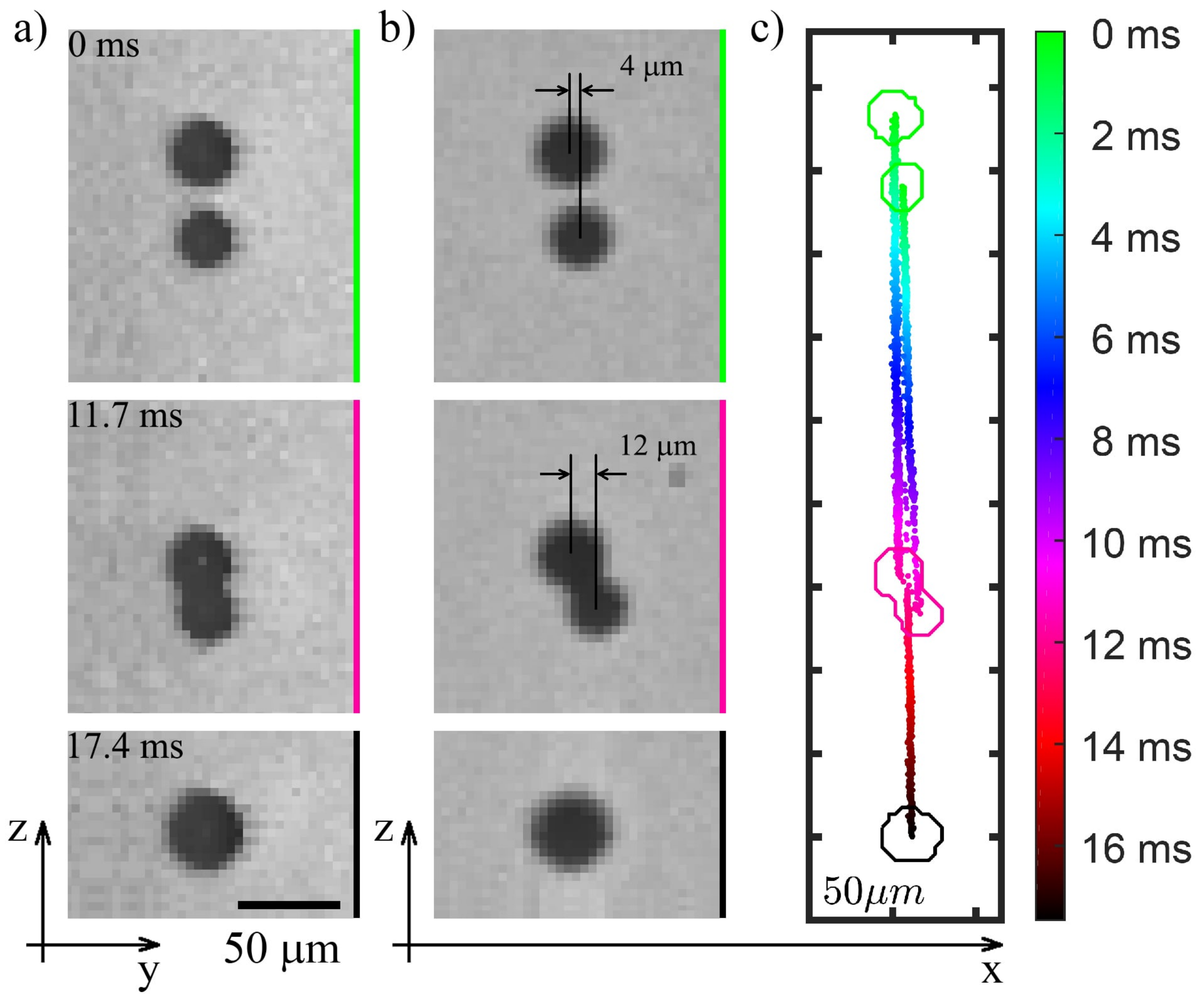}
\caption{Snapshots of two droplets that interact without the influence of laser light resulting in coalescence as seen on (a) the $y$-$z$ plane and (b) the $x$-$z$ plane. Panel (c) shows the trajectories of the droplets rendered from the images featured in panel (b). Adjacent tick marks on both axes indicate spatial separation of \SI{50}{\micro\metre}. Color bar indicates temporal progression in milliseconds.}
\label{fig:GravitationalCoalescence}
\end{figure}

\begin{figure}[ht!]
\begin{center}
\subfigure[]{
\includegraphics[scale=0.27]{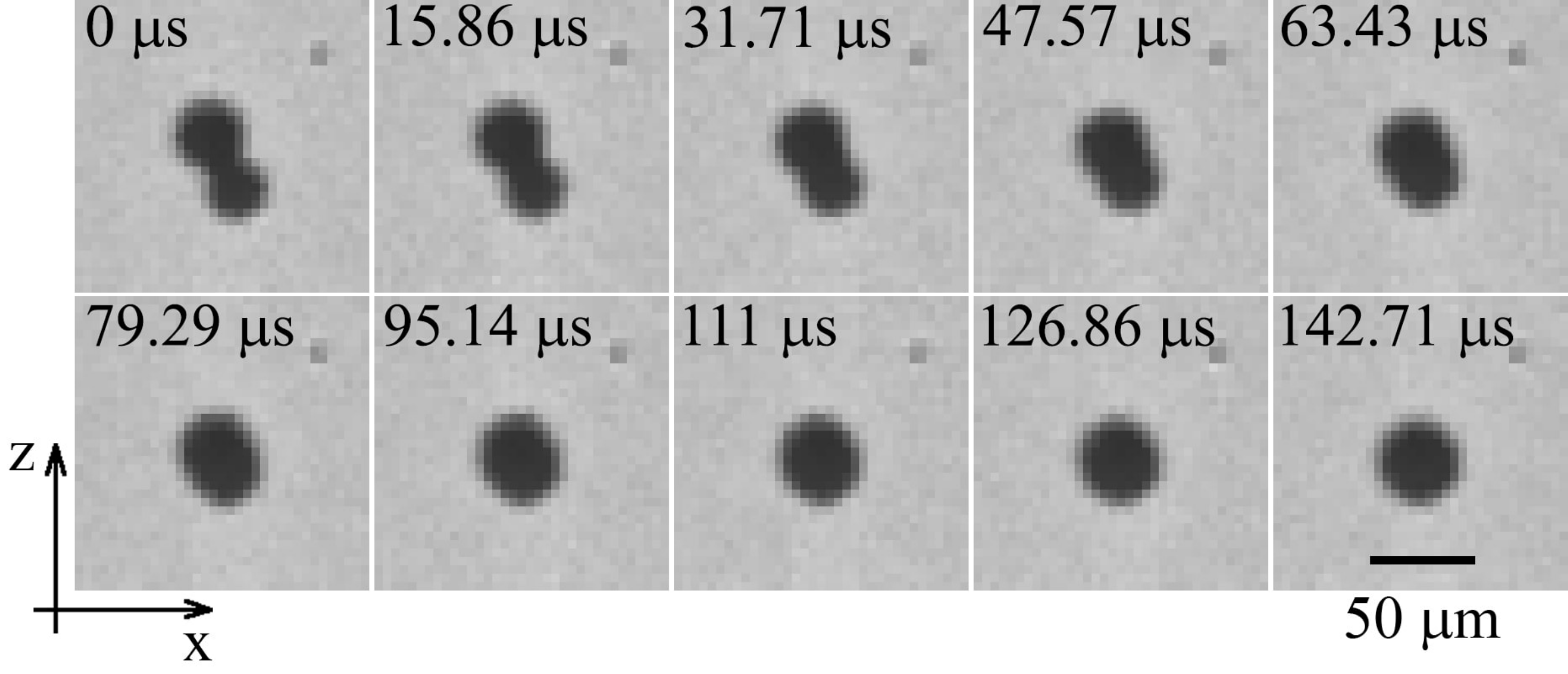}}
\subfigure[]{
\includegraphics[scale=0.27]{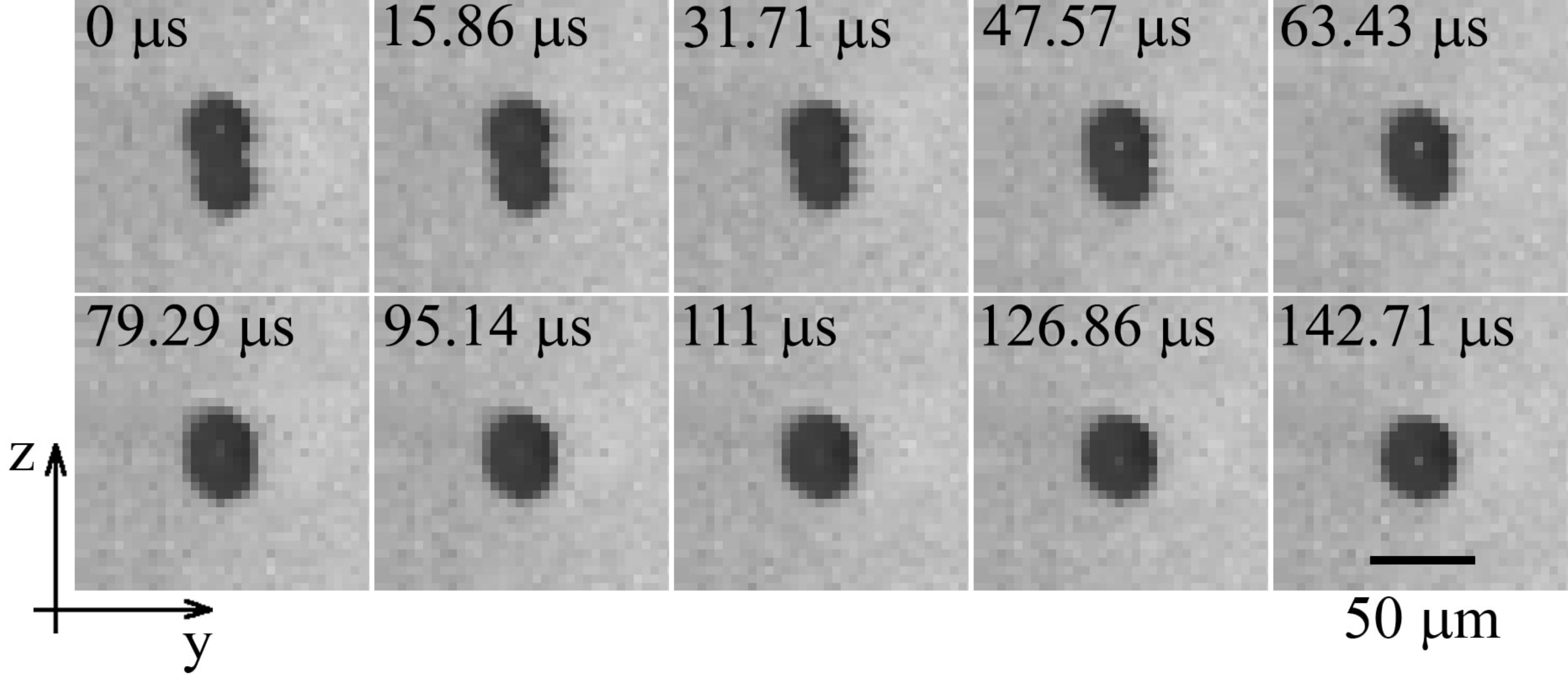}}
\caption{Frame-by-frame sequence of images showing the coalescence of two droplets sedimenting in quiescent air as seen on the (a) $x$-$z$ plane and (b) the $y$-$z$ plane.}
\label{fig:GravitationalCoalescenceProjections}
\end{center}
\end{figure}

In Fig. \ref{fig:GravitationalCoalescence} and \ref{fig:GravitationalCoalescenceProjections}, the sequences of images show the coalescence process of droplets settling under gravity and interacting without the geometrical confinement of the laser light (see \textcolor{blue}{Visualization 3}), imaged at 63~kHz.
The diameters of droplets before interaction are $29.9 \pm 2.0$ and $33.9 \pm$\SI{2.0}{\micro\metre}, and the diameter of the resultant droplet is $37.9 \pm$\SI{2.0}{\micro\metre}.
The impact parameter, calculated from the projection of droplet separations $\chi_x$ and $\chi_y$, is ${\rm B} = 0.14$, indicating a near head-on collision.
The separation between the centers of the droplets along the $x$ axis is $\chi_x =$ \SI{5.98}{\micro\metre} (0~ms in Fig. \ref{fig:GravitationalCoalescence}).
During the next $11.7$~ms, as both droplets approach each other, the larger droplet displaces the smaller droplet, thereby increasing $\chi_x$ to as much as \SI{12}{\micro\metre} prior to coalescence.
The droplets spent as much as $5.8 \pm 0.2$~ms together during which time their center-to-center separation stays less than or equal to the sum of their radii $a_1 + a_2$.
During coalescence the droplets are being pulled towards each other by capillary forces.
The center of mass of the newly formed droplet is in between those of the individual droplets.
The duration of the coalescence process is approximately \SI{150}{\micro s}.
In the accompanying movie (see \textcolor{blue}{Visualization 3}), the temporal resolution of the image acquisition system is \SI{15.86}{\micro s} from frame to frame, but for clarity the movie has been slowed down to 100~fps.

Similar to previous section, the horizontal separation between droplets increases as the droplets approach each other (Fig. \ref{fig:GravitationalCoalescence}).
But on the contrary, the droplet interaction here results in coalescence.
The main difference, between this and the previous section, is the value of the impact parameter: ${\rm B} = 0.14$ in the case of coalescence and ${\rm B} = 0.5$ in the case of non-coalescence.
A smaller impact parameter allows the  droplets to come into physical contact.
Our estimation of the collision kinetic energy (CKE) and the surface energy (SE) of colliding droplets (as discussed in \cite{low:1982} and \cite{szakall:2014}) showed that the CKE is insufficient to overcome the SE of the droplets (the CKE is 4 orders of magnitude less than the SE).
This implies that the role of the CKE is to bring the two droplets close to each other, so that other intermolecular forces could activate the coalescence process.


\section{Discussion}
\label{sec:Discussion}

In the present study we have implemented an optical trapping technique to measure the collision trajectories of free-falling droplets.
We discuss in the following a number of points that require further considerations.

\subsection{Thermal effects}
In addition to thermal fluctuations in the surrounding air, two major sources of heat are present in the experiment.
First, thermal inkjet technology uses heat, as opposed to mechanical pressure, to force the liquid out of the print head nozzle.
Second, although the absorption coefficient of glycerol is small, the focusing effect due to the spherical shape of the droplet invariably heats up the droplets and the air around it.
This may have three implications.
First, the droplets evaporate and decrease in size.
We have measured the evaporation rate of trapped droplet.
After the droplet has reached an equilibrium in the trap, the droplet size would have to reduce by as much as \SI{0.5}{\micro\metre} in diameter in order to produce a detectable change in the image.
This process happens across a time scale of approximately 10~minutes.
We found that the droplet evaporates according to the $d^2$-law \cite{frohn:2000}, that is the radius decreases in time following the relation $r^2(t) = r^2(0) - \beta \, t$, where $r(t)$ is the droplet radius at a given time $t$, $r(0)$ is the initial droplet radius and $\beta$ is the evaporation coefficient.
On the other hand, the time scale during which the droplets interact is of the order of 10~ms, which is approximately 10000 times smaller than the time scale of evaporation.
We therefore conclude that the effect of evaporation is negligible in the collision process.

Second, the temperature gradient between the droplet and the surrounding air may give rise to convective air flow and so it raised the question ``How much do the droplet trajectories deviate from those derived purely based on isothermal hydrodynamic considerations?''.
We did notice that small droplets ($a \leqslant$ \SI{2}{\micro\metre}) occasionally move upwards when released from the trap whereas large droplets ($a >$ \SI{2}{\micro\metre}) consistently fall in the same direction as gravity without any apparent convective motion.
The droplets reported in this work are typically in the size range of $a \geqslant$ \SI{15}{\micro\metre}, so we believe convective motion should be negligible.

Third, the glycerol droplet surface tension diminishes linearly with increasing temperature, albeit very weakly \cite{GPA:1963}.
The surface tension decreases by 7.6\% from a value of $6.34 \times 10^{-4}$~Nm$^{-1}$ at \SI{20}{\celsius} to $5.86 \times 10^{-4}$~Nm$^{-1}$ at \SI{90}{\celsius}.
Since the optical quality of the upper and lower traps are nearly identical, the droplets should be in nearly the same thermodynamic state upon contact, so that the coalescence mechanism is purely governed by hydrodynamical processes, and that there is no formation of thermal shock waves.
The thermal conditions of the droplet environment is certainly one aspect we can control and improve in subsequent experiments.

\subsection{Chemical and aerosol contamination}
We precisely control the volumetric mixing ratio of the water and glycerol solution in each experiment.
However, the chemical and aerosol content of the room air is not monitored.
For the individual experiments presented here, we have taken precautionary steps to reduce the probability of contamination by shielding the levitation chamber from its external environment, as described in section \ref{subsec:Chamber}, and isolating the chamber from the room air for up to a day so that undesirable contaminants may settle out of the observation volume.
Our current effort is to eliminate this uncertainty by filtering the air in a recirculation loop.

\subsection{Electrical charge}
Perhaps the largest uncertainty in connection with the hydrodynamics of the collision process is the state of the electrostatic charge on the droplets in our experiments.
Earlier experiments utilizing nozzle-based droplet generation technique have noted accumulation of charges on the droplets upon ejection from the nozzle orifice \cite{lu:2010}.
While the presence of electrical charge may be an undesirable aspect in a hydrodynamic experiment, the dynamics of charged droplets is of particular relevance to the atmosphere \cite{macgorman:1998} and protoplanetary nebula \cite{cuzzi:2001}.
In the context of our experiment, which concerns the collision process of binary droplets in a gravitational field, the presence of charge may add complexity but is not at all unfavorable.
Possible diagnostic measures to determine the amount of electrical charge on individual droplets include applying a DC or AC electric field across the droplet and measuring the displacement from its equilibrium position \cite{isaksson:2013}.

\subsection{Mixing time scale}
Upon initiation of the coalescence process, mixing occurs internally inside the newly formed droplet.
The characteristic time scale of the mixing process can be estimated using dimensional analysis \cite{tang:2012}.
Due to size dissimilarity, the capillary pressure within the small droplet exceeds that within the large droplet by an order of magnitude of $2 \, \sigma \, (1/a_2 - 1/a_1)$.
This small pressure difference drives the small droplet into the large droplet and disperses the liquid inside the newly formed droplet.
This is a fair assumption because large scale motion is most effective at transporting momentum.
The characteristic velocity of the large scale motion, $u$, can be obtained from the dimensional relation $\rho_d \, u^2 / 2 = 2 \, \sigma \, (1/a_2 - 1/a_1)$.
This motion occurs at a time scale given by $T = D / u$, where $D$ is the diameter of the newly formed droplet.
For a \SI{37.9}{\micro\metre} droplet as discussed in section \ref{subsec:GravitationalCoalescence}, the corresponding time scale is approximately \SI{30}{\micro s}.
Such a motion is within the resolution limit of our imaging system and is an interesting topic for further research.

\section{Conclusion}
\label{sec:Conclusion}
We have developed a new technique to image the trajectories of a pair of micron-sized droplets settling and colliding under gravity.
Unlike previous studies with optical tweezers, in our experiment the droplets interact without the confinement of the optical trap.
We have full control over the initial conditions of the collision process, namely the impact parameter, the size ratio, and the chemical compositions of the droplets.

For large impact parameter (${\rm B} = 0.5$), we have observed that the approaching droplets repel each other from their settling trajectories.
For small impact parameter (${\rm B} = 0.14$), the collision results in permanent coalescence.
To the best of our knowledge this is the first attempt to probe interactions of micron-sized droplets under gravity with fully controllable initial conditions.
The experiment described here indicates the potential of the technique for studying the important problem of droplet growth by gravitational collision of cloud droplets.
In more sophisticated experiments, the technique could be used to obtain fluorescence spectra of colliding droplets \cite{ivanov:2016}, and to probe collision between electrically charged droplets, post-collision dissipative phenomena and rotational dynamics of spinning droplets.
Progress in these areas of study will advance our understanding of the strong interactions between turbulence and the microphysical processes in clouds \cite{bodenschatz:2010}.

\section*{Funding}
Knut and Alice Wallenberg Foundation (Dnr. KAW 2014.0048) ``Bottlenecks for particle growth in turbulent aerosols''. The Swedish Institute scholarship through the Visby Programme.

\section*{Acknowledgments}
The authors thank J. Wettlaufer for valuable discussions.

\end{document}